\documentclass[11pt,a4paper]{article}
\usepackage{amssymb} \usepackage{amsmath} \usepackage{graphicx}
\usepackage{slashed}
\usepackage{epsfig,latexsym,color,xcolor}
\usepackage{ulem}
\usepackage{amstext}    
\usepackage{array} 
\usepackage{amsmath}
\begin{document}
\def\Giulia{\bf\color{red}}
\def\bef{\begin{figure}}
\def\eef{\end{figure}}
\newcommand{\ans}{ansatz }
\newcommand{\be}[1]{\begin{equation}\label{#1}}
\newcommand{\beq}{\begin{equation}}
\newcommand{\ee}{\end{equation}}
\newcommand{\beqn}[1]{\begin{eqnarray}\label{#1}}
\newcommand{\eeqn}{\end{eqnarray}}
\newcommand{\bd}{\begin{displaymath}}
\newcommand{\ed}{\end{displaymath}}
\newcommand{\mat}[4]{\left(\begin{array}{cc}{#1}&{#2}\\{#3}&{#4}
\end{array}\right)}
\newcommand{\matr}[9]{\left(\begin{array}{ccc}{#1}&{#2}&{#3}\\
{#4}&{#5}&{#6}\\{#7}&{#8}&{#9}\end{array}\right)}
\newcommand{\matrr}[6]{\left(\begin{array}{cc}{#1}&{#2}\\
{#3}&{#4}\\{#5}&{#6}\end{array}\right)}
\newcommand{\cvb}[3]{#1^{#2}_{#3}}
\def\lsim{\raise0.3ex\hbox{$\;<$\kern-0.75em\raise-1.1ex
e\hbox{$\sim\;$}}}
\def\gsim{\raise0.3ex\hbox{$\;>$\kern-0.75em\raise-1.1ex
\hbox{$\sim\;$}}}
\def\abs#1{\left| #1\right|}
\def\simlt{\mathrel{\lower2.5pt\vbox{\lineskip=0pt\baselineskip=0pt
           \hbox{$<$}\hbox{$\sim$}}}}
\def\simgt{\mathrel{\lower2.5pt\vbox{\lineskip=0pt\baselineskip=0pt
           \hbox{$>$}\hbox{$\sim$}}}}
\def\unity{{\hbox{1\kern-.8mm l}}}
\newcommand{\eps}{\varepsilon}
\def\ep{\epsilon}
\def\ga{\gamma}
\def\Ga{\Gamma}
\def\om{\omega}
\def\omp{{\omega^\prime}}
\def\Om{\Omega}
\def\la{\lambda}
\def\La{\Lambda}
\def\al{\alpha}
\def\beq{\begin{equation}}
\def\eeq{\end{equation}}
\newcommand{\ov}{\overline}
\renewcommand{\to}{\rightarrow}
\renewcommand{\vec}[1]{\mathbf{#1}}
\newcommand{\vect}[1]{\mbox{\boldmath$#1$}}
\def\tm{{\widetilde{m}}}
\def\mcirc{{\stackrel{o}{m}}}
\newcommand{\Dm}{\Delta m}
\newcommand{\dm}{\varepsilon}
\newcommand{\tanb}{\tan\beta}
\newcommand{\nbar}{\tilde{n}}
\newcommand\PM[1]{\begin{pmatrix}#1\end{pmatrix}}
\newcommand{\up}{\uparrow}
\newcommand{\down}{\downarrow}
\newcommand{\refs}[2]{eqs.~(\ref{#1})-(\ref{#2})}
\def\omE{\omega_{\rm Ter}}
\newcommand{\eqn}[1]{eq.~(\ref{#1})}
%

\newcommand{\DSUSY}{{SUSY \hspace{-9.4pt} \slash}\;}
\newcommand{\DCP}{{CP \hspace{-7.4pt} \slash}\;}
\newcommand{\mc}{\mathcal}
\newcommand{\gr}{\mathbf}
\renewcommand{\to}{\rightarrow}
\newcommand{\gtc}{\mathfrak}
\newcommand{\wh}{\widehat}
\newcommand{\br}{\langle}
\newcommand{\kt}{\rangle}


\def\lsim{\mathrel{\mathop  {\hbox{\lower0.5ex\hbox{$\sim$}
\kern-0.8em\lower-0.7ex\hbox{$<$}}}}}
\def\gsim{\mathrel{\mathop  {\hbox{\lower0.5ex\hbox{$\sim$}
\kern-0.8em\lower-0.7ex\hbox{$>$}}}}}

\def\nn{\\  \nonumber}
\def\de{\partial}
\def\brf{{\mathbf f}}
\def\bbf{\bar{\bf f}}
\def\bF{{\bf F}}
\def\bbF{\bar{\bf F}}
\def\bA{{\mathbf A}}
\def\bB{{\mathbf B}}
\def\bG{{\mathbf G}}
\def\bI{{\mathbf I}}
\def\bM{{\mathbf M}}
\def\bY{{\mathbf Y}}
\def\bX{{\mathbf X}}
\def\bS{{\mathbf S}}
\def\bb{{\mathbf b}}
\def\bh{{\mathbf h}}
\def\bg{{\mathbf g}}
\def\bla{{\mathbf \la}}
\def\bmu{\mathbf m }
\def\by{{\mathbf y}}
\def\bmu{\mbox{\boldmath $\mu$} }
\def\bsig{\mbox{\boldmath $\sigma$} }
\def\bunity{{\mathbf 1}}
\def\cA{{\cal A}}
\def\cB{{\cal B}}
\def\cC{{\cal C}}
\def\cD{{\cal D}}
\def\cF{{\cal F}}
\def\cG{{\cal G}}
\def\cH{{\cal H}}
\def\cI{{\cal I}}
\def\cL{{\cal L}}
\def\cN{{\cal N}}
\def\cM{{\cal M}}
\def\cO{{\cal O}}
\def\cR{{\cal R}}
\def\cS{{\cal S}}
\def\cT{{\cal T}}
\def\eV{{\rm eV}}

\numberwithin{equation}{section}

\vspace{6mm}

\large
 \begin{center}
 {\Large \bf  
 A modern guide to {\rm $\theta$}-Poincar\'e}

 \end{center}

 \vspace{0.1cm}



\begin{center}
{\large Andrea Addazi \& Antonino Marcian\`o}\footnote{E-mail: \, andrea.addazi@qq.com, marciano@fudan.edu.cn}
\\
{\it Department of Physics \& Center for Field Theory and Particle Physics, Fudan University, 200433 Shanghai, China}
\end{center}

\vspace{1cm}
\begin{abstract}
\large
\noindent 
\noindent 
Motivated by the recent interest in underground experiments phenomenology (see Ref.~\cite{Addazi:2017bbg}), we review the main aspects of one specific non-commutative space-time model, based on the Groenewold-Moyal plane algebra, the $\theta$-Poincar\'e space-time. In the $\theta$-Poincar\'e scenario, the Lorentz co-algebra is deformed introducing a non-commutativity of space-time coordinates. In such a theory, a new quantum field theory in non-commutative space-time can be reformulated. Tackling on several conceptual misunderstanding and technical mistakes in the literature, we will focus on several issues such: $i)$ the construction of fields theories in $\theta$-Poincar\'e; $ii)$ the unitarity of the S-matrix; $iii)$ the violation of locality, $iv)$ the violation of the spin statistic theorem and the Pauli principle; $v)$ the observables for underground experiments. 
\end{abstract}

\baselineskip = 20pt

\section{Introduction \& Conclusions}


It is commonly retained that no any bounds to quantum gravity theories may be inferred from current or next future experiments. The naive motivation is that the energy-scales reached by current experiments, such as collider physics, are far below the characteristic scale of quantum gravity, the Planck scale, for several orders of magnitude. For example, it is worth to remind that the Large Hadron Collider (LHC) can probe energy scales of about $1-10\, {\rm TeV}$ or so, which is 15th $-$ 16th order of magnitude down to the Planck scale ($10^{16}\,{\rm TeV}$). This is the main argument inspiring a certain pessimistic approach to quantum gravity phenomenology, which is considered more a {\it metaphysics chimera} or even an {\it oxymoron} than a serious possibility. There are several popular scenarios suggested in literature leaving a tiny hope of detecting quantum gravity effects in high energy colliders. These scenarios are based on the long-standing idea of having large extra dimensions, such as in Arkani-Hamed/Dvali/Dimopoulos (ADD) \cite{ArkaniHamed:1998rs} and Randall/Sundrum (RS) models \cite{Randall:1999ee}. Such models are often adduced as a possible solution to the hierarchy problem of the Higgs mass. However, in this context, the hierarchy problem seems to be only rephrased as a problem of stabilizing the extra dimension sizes, i.e. their related moduli as the radion field. Furthermore, a quantum gravity scale so low may also generate dangerous effective operators, destabilizing baryons \cite{Adams:2000za}. Such a problem may be solved by introducing extra gauge or flavor symmetries \cite{Berezhiani:1998wt}, turning the model in a more complicated and baroque. 

Related to the last argument, it is possible that a large class of quantum gravity models may  induce highly suppressed effective operators beyond the Standard Model of particle physics. An open testable possibility, considered in our analysis in Ref.~\cite{Addazi:2017bbg}, consists in searching for exotic transitions in nuclei or atoms which violates the Pauli exclusion principle. Such a possible discovery has the terrific potentiality to change our same conception of space and time. In fact, the Pauli Principle is directly related to the Spin Statistic theorem of the Standard Model of particle physics. In turn the Spin Statistic theorem is related to the same property of the Minkowski's space-time, causality, locality and the Poincar\'e symmetry group. The detection of Pauli Exclusion Principle Violations (PEPV) may open the intriguing possibility of detecting indirect quantum gravity smoking guns in underground experiments of rare processes physics. A quantum gravity model that is predicting Pauli Exclusion Principle violating transitions is the $\theta$-Poincar\'e space-time. 

\subsection*{$\theta$-Poincar\'e in a nutshel}

This is one of the most popular scenario of non-commutative space-time. It is based on a deformation of the Poincar\'e symmetry, and it entails a dual formulation in terms of non-commutative space-time coordinates. The idea of a non-commutative space-time was proposed by the same Heisenberg, as a simple but conceptually revolutionary extension of his indetermination principle \cite{Heisenberg}. However, in '47, Snyder \cite{Snyder} and Yang \cite{Yang} were the first to seriously propose it in published papers. 

The $\theta$-Poincar\'e co-algebra can be obtained from the Poincar\'e algebra thanks to a mathematical map, known as the Groenewold-Moyal (GM) map
\cite{Majid:1996kd,Oeckl:2000eg,Chaichian:2004za,Aschieri:2005yw,Balachandran:2004rq,Balachandran:2005eb,Chaichian:2002vw,Balachandran:2006ib,AlvarezGaume:2001ka,Greenberg:2005jq,AlvarezGaume:2003mb,Haque:2007rb}. On the dual space-time formulation, deformation of the co-algebra is encoded in the product of generic functions $f(x)$ and $g(x)$ of the commutative coordinates $x\equiv (x_{1},...x_{\rm N}) \in \mathbb{R}^{\rm N}$. The deformation of the co-products introduced by the GM map then acts a ``$\star$ product" between functions. For instance, considering two generic functions, the $\star$ product casts 

\be{fg}
f  \star g= f \ \exp\Big(\frac{\imath}{2}\overleftarrow{\partial}_{\mu}\theta^{\mu\nu}\overrightarrow{\partial}_{\nu} \Big)  \ g
\ee
where 
$$\theta^{\mu \nu}=-\theta^{\nu \mu}$$ 
is a antisymmetric constant tensor in space and time. In the limit in which all the components $\theta_{\mu\nu}\rightarrow 0$, the product in Eq.~(\ref{fg}) trivializes to the standard commutative multiplication rule. On the other hand, because of the antisymmetry of the $\theta$-matrix, $f \star g\neq g \star f$. 

In the $\theta$-Poincar\'e framework, a consistent quantum field theory can be obtained. 
Non-commutative coordinates might be viewed as quantum operators
\begin{equation}
\label{xxmu}
\hat{x}^{\mu}(x)=x^{\mu} 
\end{equation}
endowed with the $\star$-product in Eq.~\eqref{fg}.
This latter, once applied to the coordinate-operator, implies 
 \begin{equation}
 \label{jkll}
 (\hat{x}^{\mu}\star \hat{x}^{\nu}-\hat{x}^{\nu}\star \hat{x}^{\mu})=[\hat{x}^{\mu},\hat{x}^{\nu}]_{\star}=\imath \theta^{\mu\nu}\, ,
\end{equation}
which is exactly the non-commutativity proposed by Heisenberg. 

The very same product rule must be applied to every quantum field theory operators:
creation/annihilation particle operators and every fields (electro-weak, chromo-strong and Higgs fields). In other words, a deformed version of the Standard Model of particle physics may be obtained as a Groenewold-Moyal Standard Model (GSSM). In GSSM, one can easily obtain all GM Feynman diagrams from the standard ones. However, most of the amplitudes are corrected by harmonic functions, which are dependent on the particles four-momenta. In the limit of $\theta\rightarrow 0$, at three level, all amplitudes converge to Standard Model results.


Before moving into the details of the $\theta$-Poincar\'e model, it is worth spelling a simplified and heuristic picture of its deformation. This may give the chance for introducing the mathematical framework in a more accessible way to experimentalists and phenomenologists rather than theorists devoting their investigation on quantum gravity. 
At this purpose, a first question is: how are the generators of the Poincar\'e algebra deformed by the GM map? Surprisingly, space-time translations $x^{\mu}\rightarrow x^{\mu}+a^{\mu}$ 
are untouched. By applying GM on translation operator $p_{\mu}$, in turn applied to fields, we obtain exactly the same result of Standard Poincar\'e case: 
$${\bf GM}:\, {\rm translation} \rightarrow {\rm translation}\, . $$
The same ``triviality" is not fund for the case of the Lorentz algebra --- namely $\mathfrak{so}(3,1)$ --- generators: 
$${\bf GM}:\,\mathfrak{so}(3,1)\rightarrow {\rm noncommutative \ dual}\,\,\,{\rm ``deformed''}\,\,\, \mathfrak{so}(3,1)\,. $$
In particular, Lorentz rotations and boost transformations $x_{\mu}=\Lambda_{\mu}^{\ \nu} \, x_{\nu}$ are deformed with a complicated combinations of four-momenta dependent phases that come from the Moyal product Eq.~(\ref{fg}). As aforementioned, not only space-time generators, but also quantum operators related to particle fields are deformed as follows: 
$${\bf GM}: \, {\rm (creation/annihilation\, ops.)} \rightarrow {\rm (GM-phase)(creation/annihilation\, ops.)\rm}\,,$$
$${\bf GM}:\,{\rm (fields)} \rightarrow {\rm (GM-phase)(fields)}\,,$$
$${\bf GM}:\, {\rm N-field}\,\,{\rm interactions}\rightarrow {\rm (GM-phase)^{N}(creation/annihilation\, ops.)^{N}}\,.$$
The GM-map also provides a non-ambiguous relation among the quantization procedure of fields in $\theta$-Poincar\'e and the standard second quantization in the Standard Model. To fully appreciate how powerful is this property for the $\theta$-Poincar\'e framework, let us mention that there are other non-commutative QFT models, such as the $\kappa$-Poincar\'e one, which are plagued by several ambiguities in the quantization procedure
 \cite{Arzano:2007ef,AmelinoCamelia:2007uy,AmelinoCamelia:2007vj,AmelinoCamelia:2007wk,AmelinoCamelia:2007rn,Marciano:2008tva,Marciano:2010jm,Agostini:2006nc,Arzano:2007gr,AmelinoCamelia:2007zzb,Marciano:2010gq,AmelinoCamelia:2010zf,AmelinoCamelia:2012it}. \\

It is also worth noting that the GM factors, encoding four-momentum dependent complex exponentials, introduce in the theory an infinite tower of new higher derivative terms. In other words, the GM co-product rule provides an unambiguous map from a local QFT into a new Lorentz deformed higher derivatives QFT. In this review, our main purpose is to provide a clear explanation of many crucial issues, such as unitarity, causality, locality and $\mathcal{CPT}$ symmetry in $\theta$-Poncar\'e. These were sources of a huge confusion in literature, with often too naive considerations concerning intimate interconnections among them all. 

\subsection*{Myths and facts of $\theta$-Poincar\'e}

First of all, $\theta$-Poincar\'e does not lie into the class of theories explicitly violating the Lorentz symmetry: the Lorentz symmetry principle is substituted by a new symmetry algebra. In this sense, it is more precise to consider $\theta$-Poincar\'e as a deformation of the Lorentz symmetry despite of a violation. An explicit violation of the Lorentz symmetry introduces an infinite tower of new interaction operators, which renders the well behaving standard model a non-renormalizable not-unitary ``nightmare" theory \cite{Collins:2004bp}. Even introducing very fine-tuned Lorentz violating operators, it may be very dangerous since not technically natural fine-tunable ---  with coupling exploding polynomially with the energy. As we will see, $\theta$-Poincar\'e looks a ``much milder beast". 

In $\theta$-Poincar\'e also a curious fact occurs: the theory is explicitly unitary a three level, even if the S-matrix does not commute with $\mathcal{CPT}$. Once again, it would be more precise to talk about a deformation of the standard $\mathcal{CPT}$. It is worth noticing that $\mathcal{CPT}$-phases completely disappear if the ``electric-like" components of $\theta_{\mu\nu}$ are null, i.e. $\theta_{0i}=0$ but $\theta_{ij}\neq 0$. To easily understand why the theory remains unitary, let us mention that the GM phases introduced into the fields, and propagating through all the Hamiltonian interaction operators, do not spoil the hermitianity of the Hamiltonian. In other words, no ``badly behaving'' complex phases are introduced in the S-matrix. 

On the other hand, locality and microcausality are explicitly violated in these theories. It is worth noting that the very same structure of noncommutative space-time is intimately non-local and acausal at microscopic lengths. This does not mean that large acausal interactions are introduced in the Standard Model. A simple criterion of microcausality in quantum field theory was suggested by Bogolubov and Shirkov (BS) in their quantum field theory textbook \cite{BS}. According to BS, the microcausality condition can be directly reformulated in terms of the S-matrix operator. In our review, we will show that the microcausality criterion is violated in $\theta$-Poincar\'e, even at tree-level. However, such a violation may be not-relevant if the UV quantum gravity cut-off coincides with the non-commutativity scale. 

All these considerations suddenly rise further questions about the Spin-Statistic theorem,
formulated by Pauli in Ref.~\cite{Pauli:1940zz}. The Spin-statistic theorem was conceived assuming locality, causality, Poincar\'e and $\mathcal{CPT}$ invariance. By relaxing these principles, one could expect an evasion of the Pauli theorem. And indeed this is exactly what happens for $\theta$-Poincar\'e. As we will see, this leads to the most surprising and phenomenologically interesting aspect of this theory: the sharp prediction of Pauli exclusion principle violating transitions. 

In our review, in order to show how treatable are tree-level calculations in $\theta$-Poincar\'e, we will show some examples in standard QED processes: electron-muon and electron-electron scatterings. Remarkably, in the electron-muon scattering, all $\mathcal{CPT}$ violating phases completely disappear --- incidentally, we emphasize that such an interesting fact was never remarked in literature. On the other hand, in the electron-electron scattering a $\mathcal{CPT}$ violating phase appears as a harmonic function with particle momenta. Such a $\mathcal{CPTV}$ phase emerges as a decoherence effect of s- and t-channel interferences. Once again, this is a manifestation of the Spin-Statistic theorem evasion. Nevertheless, such a phase never leads to a violation of the unitarity bound and, at low energies, it simply provides a tiny over-modulation on the Standard Model differential cross-section.  At three level everything is calculable, controllable and theoretically self-consistent. 

\subsection*{Why $\theta$-Poincar\'e?}

It seems that most of the models that are candidates of quantum gravity ``like" $\theta$-Poincar\'e. Within the context of string theory, Seiberg and Witten have shown that
a $\theta$-Poincar\'e QFT can emerge as a low energy limit of String Theory coupled with a $B$-form background field \cite{Seiberg:1999vs}. As it is well known, open and closed bosonic strings $X^{M}(\sigma,\tau)$ can be considered as maps from the string $1+1$ dimensional conformal invariant world-sheet to a $M=9+1$-dimensional space-time. In other words, $X$-fields have a double interpretation: they are ten scalar fields in the worldsheet and a one vector-field in the target extra dimensional space-time. On the other hand, antisymmetric tensor fields $B_{MN}$ emerges in every possible closed string theory, universally, together with the graviton and the dilaton fields. It is then likely natural to consider a coupling such as $B \cdot (\partial X\partial X)$. A possible issue may concern coupling a symmetric tensor combination $(\partial X\partial X)$ to an antisymmetric tensor $B$. However, from the world-sheet side, one can introduce an extra $\epsilon_{ab}$ anti-symmetrizer, and then consider $\epsilon_{ab}B^{MN}\partial^{a}X_{M}\partial^{b}X_{N}$. 
As we will review later on, such a coupling is responsible for the generation of an effective 
non-commutativity affecting bosonic string $X$-fields. 

On the other hand, an effective non-commutativity may also emerge within the context of Loop Quantum Gravity, as an effective mesoscopic limit of the spin foam \cite{AmelinoCamelia:2003xp,Freidel:2005me,Cianfrani:2016ogm,Amelino-Camelia:2016gfx,NClimLQG,BRAM,Brahma:2017yza}. The theory retains a lattice smeared version  
of the Lorentz invariance, but at intermediate scale rainbow behavior of the metric \cite{Lewandowski:2017cvz} and space-time symmetries modification appear \cite{Gambini:1998it, Alfaro:1999wd}. What seems to be remarkable, is that this feature pertains not only the $2+1$-dimensional theory, but also extends to $3+1$-dimensions, standing as a residual property of the discreteness space (and space-time) in at microscopic (Planckian) scales. Furthermore, it turns out to be connected to an intrinsic feature of the space-time background, once matter fields are integrated out, or anyway matter degrees of freedom are not taken into account. 

Behind the common effective non-commutativity that emerge both in String Theory and Loop Quantum Gravity, there may be a hidden duality among the two theories. The two theories appear so different that such a statement may sound unnatural and crazy (perhaps,  crazy enough to hold some truth). Very recently, we singled out a series of correspondences among spin foam holonomies and space-like branes, that we dubbed Hilbert-duality, or ${\bf H}$-duality \cite{Addazi:2017qwt,Addazi:2018cyn}. It turned out that a self-dual reformulation of gravity {\it \`a la} LQG can be cast from the compactification of topological M-theory.

Finally, non-commutative quantum gravity was also proposed as a self-consistent candidate of UV completed quantum gravity, avoiding the singularity problem as an effect of the non-commutative indetermination principle. In particular, there were a series of attempts of Connes, Chamsedinne, Mukhanov et al to unify quantum gravity and Grand Unification theories in a non-commutative framework \cite{Chamseddine:2014nxa,Connes:2017oxm,Chamseddine:2010ud,Chamseddine:2013rta}. However, full conclusive concrete calculations in the case of quantum black hole or Big Bang were never shown in the literature, and it seems that the status of the field, concretely, is still stuck to mathematical formalities.

\subsection*{Into the wild: UV/IR mixings}

A relevant issue in $\theta$-Poincar\'e is the extremely complicated behaviour of radiative corrections. Switching on $\hbar$ in loop corrections, something very exotic happens: it seems that terms behaving like $1/(\theta^{2}p^{2})$ appear. These 
can be interpreted as a cut-off energy dependence, as $\Lambda_{\rm eff}=\Lambda_{0}+C/(\theta^{2} p^{2})$, where $C$ is a model dependent factor. Such an occurrence represents a very dangerous phenomenon: it seems that in the IR limit $p\rightarrow 0$, the UV cutoff diverges --- in other words, the commutative limit $\theta \rightarrow 0$ diverges. The phenomenon is then called UV/IR mixing and it was probably the main motivation for the decrease of interest in the theoretical community on $\theta$-Poincar\'e. 
For example, we will show that the photon propagator gets a divergent correction which leads to a disastrous ghost-like correction to its propagator, apparently leading to super-luminality. However, contrary to the claim done by several authors in literature, it is still not clear if these divergences really occur in the full theory. For instance, it is possible that radiative IR divergences are cancelled by considering the emission of real radiation from photons: if the photon dynamically gets a superluminal behaviour, it should start to emit Cherenkov's radiation, since its velocity is higher than the space-time medium speed of light
  \cite{Joshi:1986tr,Recami:1984nc,Joshi:1986tr,Folman:1995jq,Dawe:1998dr}. Photons have indeed non-abelian self-interaction terms in $\theta$-Poincar\'e that may provide a radiation emission portal mechanism. The emission of Cherenkov's radiation was studied in the case of charged tachyon particles, but never within the context of $\theta$-Poincar\'e. If the effect is not suppressed by some possible subtleties, a $O(1/p^{2}\theta^{2})$ is expected as well. In other words, our "message" about UV/IR divergences is that, before putting the word "end" to the theory, one should show a detailed calculation of competitive IR processes from both radiative and real photon insertions.

\vspace{0.5cm}

\noindent

Finally we will arrive to the {\it vexata quaestio:}

\subsection*{Is $\theta$-Poincar\'e testable?}

The answer is surprisingly not only yes, as we shown in Ref.~\cite{Addazi:2017bbg}, but even more surprisingly, the {\it democratic} implementation of $\theta$-Poincar\'e models is already ruled out by several underground experiments' data. The falsification of any model is here provided by existent very tight constraints on the Pauli exclusion principle violations (PEPV). It urges now a definition of the distinction among two possible classes of $\theta$-Poincar\'e models: the {\it democratic} and the {\it despotic} cases. The {\it democratic} $\theta$-Poincar\'e models assume that all the Standard Model fields propagate in the non-commutative space-time background with the same coupling. Such a scenario seems to be the most natural case. However, within the context of string theory, such a scenario seems to be the most fine-tuned case. As mentioned before, in string theory, non-commutativity emerges as an effect of the B-field coupling with strings. Indeed, there is no any motivation of why the couplings of the B-field with all the strings must be of the same magnitude: in analogy with the electric or the magnetic field, the charges with respect to the B-background can be different from one another. This is what we mean for {\it despotic} scenario: some strings may propagate in a effectively different space-time background. 
 
As we have discussed in our work (Ref.\cite{Addazi:2017bbg}), $\theta$-Poincar\'e can induce very tiny but testable Pauli forbidden transitions. Contrary to effective PEP violating 
models proposed in Refs.~\cite{Messiah:1900zz,Greenberg:1963kk,Gentile,Green:1952kp,Ignatiev:1987zd,Gavrin:1988nu,PEP1,PEP2,PEP3}, such transitions are: i) energy dependent from the particular PEPV process considered; ii) suppressed with the non-commutative energy scale.

In this review, we will include a detailed calculation of PEPV atomic/nuclear level transitions
induced by $\theta$-Poincar\'e. Our results show that underground experiments can rule out  $\theta$-Poincar\'e models up to non-commutative length scales beyond the Planck scale $10^{-35}\, {\rm meters}$. In other words, rare processes in nuclear and atomic physics can literally be used as an indirect probe of the same structure of space and time. 

Before addressing the details of the analysis, we may anticipate that BOREXINO, KAMIOKANDE and DAMA still exclude $\theta$-Poincar\'e in the hadronic strong sector 
beyond the Planck scale! This seems to be a phenomenological tombstone 
for the democratic $\theta$-Poincar\'e models. However, despotic models are still alive. This highly motivates measures from other kind of underground detectors. For example, an experiment such as VIP provides a measure of Pauli violating transitions in the atomic electromagnetic sector. In our paper, we will show an estimation of constrains which can be inferred by VIP from current data: they are not enough for concluding that the despotic $\theta$-Poincar\'e scenario is excluded at the Planck scale. The next update of the experiment, VIP2, may arrive to reach enough sensitivity to probe despotic $\theta$-Poincar\'e models from a complementary channel.

\section{Fields} \label{mico}
\noindent 
We introduce the Moyal plane $\mathcal{A}_{\theta}(R^{N})$ co-algebra, starting from the  $\star$-product in Eq.~\eqref{fg}. The diffeomorphism group $\mathcal{D}(R^{N})$ acts on $\mathcal{A}_{\theta}(R^{N})$ providing its coproduct in the twisted form, namely
\be{thetaa}
\Delta_{\theta}(g)=\mathcal{F}_{\theta}^{-1}(g\otimes g)\mathcal{F}_{\theta}
\ee
where
\be{Ftheta}
\mathcal{F}_{\theta}=\exp\Big(\frac{\imath}{2}\partial_{\mu} \theta^{\mu\nu} \partial_{\nu}\Big)
\ee
with $g$ is the algebra element and $\mathcal{F}_{\theta}$ called "Drienfel'd twist". 

Moving from this definition, we can construct fields as irreducible representations of the deformed Poincar\'e group. In order to do it, we must construct fields with a consistent measure structure with respect to the deformed translations and Lorentz transformations.
We may start with a generic form of the spin-less scalar field
\be{phi}
\phi=\int \!d\mu(p)\, \tilde{\phi}(p)e_{p}\, . 
\ee

The coproduct provides a multiplication map deforming the group elements $g$ of every Standard Model symmetries as follows
\be{Deltatheta}
\Delta_{\theta}(g)=e^{\frac{\imath}{2} P_{\mu}\theta^{\mu\nu}P_{\nu}}(g \otimes g) e^{-\frac{\imath}{2} P_{\mu}\theta^{\mu\nu}P_{\nu}}=\hat{F}^{-1}_{\theta}(g\otimes g)\hat{F}_{\theta}\,.
\ee

The action of the deformed translations on the field is trivially as follows: 
\be{Delta}
\Delta_{\theta}(e^{\imath P\cdot a})(\tilde{\phi}(p)\otimes \tilde{\chi}(q))=e^{\imath(p+q)\cdot a}\tilde{\phi}(p)\tilde{\chi}(q)\, . 
\ee
However, twisted Lorentz transformations acting on the fields are deformed in the following way: 
\be{twisted}
\Delta_{\theta}(\tilde{\phi}(p)\otimes \tilde{\chi)(q)}=\tilde{F}_{\theta}^{1}(\Lambda^{-1}p,\Lambda^{-1}q)\tilde{F}_{\theta}(p,q)\tilde{\phi}(\Lambda^{-1}p)\tilde{\chi}(\Lambda^{-1}q)\, , 
\ee

where 
\be{Ftheta}
\tilde{F}_{\theta}(p,q)=e^{-\frac{\imath}{2}p_{\mu}\theta^{\mu\nu} q_{\nu}}\, . 
\ee

Consistency with Eqs.~\eqref{Delta}, \eqref{twisted} requires for the fields --- in the discussions reported below we will focus on the case of $N=4$ --- the form
\be{phi}
\phi_{\theta}=\int \frac{d^{N-1}p}{2p_{0}}(a_{{\bf p}}e^{-ipx}+b^{\dagger}_{{\bf p}}e^{ipx})\,,
\ee
where $p_{0}=\sqrt{{\bf p}^{2}+m^{2}}$, and 
\be{ab}
a_{{\bf p}}=e^{-\frac{\imath}{2}p_{\mu}\theta^{\mu\nu}P_{\nu}} c_{{\bf p}}\,, \qquad  b_{{\bf p}}=e^{-\frac{\imath}{2}p_{\mu}\theta^{\mu\nu}P_{\nu}} d_{\bf p}\,,
\ee
with
\be{Pmu}
P_{\mu}=\int \frac{d^{N-1}p}{2p_{0}}(c^{\dagger}_{p}c_{p}+d^{\dagger}_{p}d_{p})p_{\mu}\, ,
\ee
and 
\be{standard}
[c_{{\bf p}},c^{\dagger}_{{\bf q}}]=[d_{{\bf p}},d^{\dagger}_{{\bf q}}]=2p_{0}\, \delta^{3}({\bf p}-{\bf q})\, . 
\ee
This construction can be generalized to generic spin fields. A direct consequence of Eq.~\eqref{phi} is a series of deformed creation/annihilation relations, as follows:  
\be{kakak}
a^{(s_{1})}(p_{1})a^{(s_{2})}(p_{2})=e^{\imath p_{1}\wedge p_{2}}a^{(s_{2})}(p_{2})a^{(s_{1})}(p_{1})\,,
\ee
\be{kakaks}
a^{(s_{1})\dagger}(p_{1})a^{(s_{2})\dagger}(p_{2})= e^{\imath p_{1}\wedge p_{2}}a^{(s_{2})\dagger}(p_{2})a^{(s_{1})\dagger}(p_{1})\,,
\ee
\be{iiiaaii}
a^{(s_{1})}(p_{1})a^{(s_{2})\dagger}(p_{2})=e^{-\imath p_{1}\wedge p_{2}}a^{(s_{2})\dagger}(p_{2})a^{(s_{1})}(p_{1})+2p_{10}\delta(\vect{p}_{1}-\vect{p}_{2})\, . 
\ee

The $\star$-product between the fields, namely
\be{partialal}
(\phi_{\theta}\star \phi_{\theta})(x)=\phi_{\theta}(x) e^{\frac{\imath}{2}\overleftarrow{\partial}_{\mu}\theta^{\mu\nu} \overrightarrow{\partial}_{\nu}}\phi_{\theta}(y)\Big|_{y=x}\, ,
\ee
will enter every interaction Hamiltonian, and can be denoted in terms of $\phi_{\theta}$, representing a ``twisted field''. Free terms in the Hamiltonian, which are quadratic in the fields, will be unaffected by the star-product, as a property of the cyclic integration measure that is usually taken into account for $\theta$-Poincar\'e models \cite{Pachol:2015qia}.

The twisted field $\phi_{\theta}$ is related to untwisted standard one as follows: 
\be{phih}
\phi_{\theta}=\phi_{0}e^{\frac{\imath}{2}\overleftarrow{\partial}_{\mu}\theta^{\mu\nu} P_{\nu}}\, . 
\ee
The expression is valid for both bosonic and fermionic fields, of arbitrary spin. 






\section{CPT}

In this section, we analyze $\mathcal{C,T}$ and $\mathcal{P}$ transformations, which are now deformed by the co-product. In general, $\theta$-Poincar\'e is not invariant under $\mathcal{CPT}$. 

The coproduct for the charge operator in the twisted case is the same one as in the standard case. This is simply because the charge conjugation operator commutes with $P_{\mu}$. Therefore, we can write
\be{DeltaCC}
\Delta_{\theta}(\mathcal{C})=\Delta_{0}(\mathcal{C})=\mathcal{C}\otimes \mathcal{C}\, , 
\ee
but
\be{DeltaPP}
\Delta_{0}(\mathcal{P})=\mathcal{P}\otimes \mathcal{P}\,,
\ee
while
\be{DeltaTT}
\Delta_{0}(\mathcal{T})=\mathcal{T}\otimes \mathcal{T}\, .
\ee
The coproduct in Eq.~\eqref{DeltaPP} is not compatible with the $\star$-product: parity does not induce an automorphism of $\mathcal{A}_{\theta}$. Differently, time-reversal induces an automorphism of $\mathcal{A}_{\theta}$ for any $\theta^{\mu\nu}$. Considering the very definition of space-time non-commutativity, Eq.~\eqref{jkll}, it follows that $\mathcal{T}: [\hat{x}_0,\hat{x}_i] \rightarrow -  [\hat{x}_0,\hat{x}_i] $. At the same time --- see e.g. Ref.~\cite{Balachandran:2004rq} ---  because of anti-linearity, one finds that $\mathcal{T}: \theta_{0i} \rightarrow - \theta_{0i}$, preserving the non-commutativity rule, and consequently the enveloping algebra of space-time functions generated. 

The action of the twisted coproduct in the case of $\mathcal{P}$ and $\mathcal{T}$ transformations is less trivial, and is recovered by twisting the action of the standard operator: 
\be{Deltatheta}
\Delta_{\theta}( \mathcal{P})=\mathcal{F}_{\theta}^{-1}\Delta_{0}( \mathcal{P})\mathcal{F}_{\theta}\, , 
\ee
\be{TTT}
\Delta_{\theta}(\mathcal{T})=\mathcal{F}_{\theta}^{-1}\Delta_{0}(\mathcal{T})\mathcal{F}_{\theta}\, .
\ee
The combination of the three deformed symmetries leads to 
\be{Deltatheta}
\Delta_{\theta}(\mathcal{CPT})=\mathcal{F}_{\theta}^{-1}\Delta_{0}(\mathcal{CPT})\mathcal{F}_{\theta}\, . 
\ee

We can analyze the action of $\mathcal{CPT}$ on the creation/annihilation and fields operators, respectively  
\be{CCC}
\mathcal{C}:c_{k}\rightarrow d_{k},\,\,\,\quad \mathcal{C}:a_{k}\rightarrow b_{k}\, , 
\ee
\be{CCCC}
\mathcal{C}:\phi_{\theta}\rightarrow \phi_{\theta}\, , 
\ee

\be{PPP}
\mathcal{P}:c_{{\bf k}}\rightarrow c_{-{\bf k}},\,\,\, \qquad \mathcal{P}:d_{{\bf k}}\rightarrow d_{-{\bf k}}\, , 
\ee

\be{PPPP}
\mathcal{P}:a_{{\bf k}}=a_{{\bf -k}}e^{\imath k_{0}\theta^{0i}P_{i}-\imath k_{i}\theta^{i0}P_{0}  }\,, 
\qquad 
\mathcal{P}:b_{{\bf k}}=b_{{\bf -k}}e^{\imath k_{0}\theta^{0i}P_{i}-\imath k_{i}\theta^{i0}P_{0}  }
 \, .
\ee
Introducing the alternative notation $\overleftarrow{\partial}^{\mu} \theta_{\mu\nu} P^{\nu}=\partial \wedge P$, we then immediately find that fields transform as 
\be{fieldP}
\mathcal{P}:\phi_{\theta}=(\mathcal{P} \phi_{0} \mathcal{P}^{-1})\, e^{\frac{1}{2} \overleftarrow{\partial} \wedge (P_0, -\vec{P})}\,, 
\ee
which induces an extra phase for the fields that reads  $e^{\imath k_{0}\theta^{0i}P_{i} + \imath k_{i}\theta^{i0}P_{0}}$. \\

Analogously, one may find for the action of the time reversal $\mathcal{T}$ the expressions 

\be{TTT}
\mathcal{T}:c_{{\bf k}}\rightarrow c_{-{\bf k}},\,\,\, \qquad  \qquad \mathcal{P}:d_{{\bf k}}\rightarrow d_{-{\bf k}}\,,
\ee
\be{TTTT}
\mathcal{T}:a_{{\bf k}}=a_{{\bf -k}} \, e^{- \imath k_{i}\theta^{ij}P_{j}}\,,
\qquad \qquad 
\mathcal{T}:b_{-{\bf k}}=  a_{{\bf -k}} \, e^{- \imath k_{i}\theta^{ij}P_{j}}\,.
\ee
Consequently, one finds that the action on fields is given by
\be{TTTTT}
\mathcal{T}: \phi_{\theta}=(\mathcal{T} \phi_{0} \mathcal{T}^{-1})\, e^{\frac{1}{2} \overleftarrow{\partial} \wedge (P_0, -\vec{P})}\,, 
\ee
which amounts to an extra phase in the fields representation at $\theta_{\mu\nu}\neq0$ that reads $e^{\imath k_{i}\theta^{ij}P_{j}}$.\\

These relations finally imply 
\be{CPTT}
\mathcal{CPT}:\phi_{\theta}= (\mathcal{CPT}\,\, \phi_{0} \,\, (\mathcal{CPT})^{-1})\, e^{\frac{1}{2} \overleftarrow{\partial} \wedge P}
\,,
\ee
which amounts to an extra phase in the fields representation at $\theta_{\mu\nu}\neq0$ that reads $e^{\imath k_{0}\theta_{0i}P_{i}+\imath k_{i}\theta_{i0}P_{0}}$. As a consequence, for $\theta_{0i}=0$ (no-electric-phases), the theory is automatically CPT invariant.

\subsection*{CPT symmetry and S-matrix}

\noindent
The S-matrix can be generically written in a time ordered exponential form --- having care of substituting all standard products with $\star$-products among coordinates and fields. This leads to the result 
\be{SM}
\mathcal{S}_{\theta}=T_{\star}{\rm exp}_{\star}\Big[-\imath \int d^{4}x\mathcal{H}_{I,\theta}(x)\Big]\,,
\ee
where the Hamiltonian density includes matter fields and gauge fields. Note that the time-ordering is redefined by meaning of the $\star$-product. The deformed Hamiltonian density $\mathcal{H}_{I,\theta}(x)$ is always Hermitian at tree level for every $\theta$. For example, for a generic interaction term, the expansion holds
\be{phiud}
\mathcal{H}^{n} \equiv \mathcal{H}_{1,\theta}\star \mathcal{H}_{2,\theta}\star ...\star\mathcal{H}_{n,\theta}=\mathcal{H}_{1,0}
\mathcal{H}_{2,0} \,
... \mathcal{H}_{n,0} \, e^{\frac{\imath}{2}\overleftarrow{\partial}\wedge P}
\equiv \mathcal{H}^{n\dagger}\,,
\ee
with $\mathcal{H}_{1}\equiv\mathcal{H}(x_{1})$, which implies the unitarity of the S-matrix, because of Eq.~(\ref{SM}). However, this may be not true at loop level. Interaction terms can be quantum corrected by divergent diagrams. The Hamiltonian non-trivially transforms under CPT as follows:
\be{HCPT}
(\mathcal{CPT})\mathcal{H}^{n}(\mathcal{CPT})^{-1}=\Big({\rm MOYAL}\Big)\mathcal{H}^{n}\,. 
\ee

\section{Non-Locality}
The GM products introduce a violation of microcausality, which can be formulated, in terms of interaction Hamiltonian, as follows: 
\be{OOO}
[\mathcal{H}_{\star}(x),\mathcal{H}_{\star}(y)]\neq 0,\,\,\, (x-y)^{2}<0\,. 
\ee

The microcausality principle can also be formulated directly on the S-matrix, generalizing the Bogoliubov-Shirkov (BS) method Ref.~(\cite{BS}). Let us expand the S-matrix promoting couplings to auxiliary fields: 
\be{Smatrix}
S[g]=1+\int dx_{1} g(x_{1})\star S_{1}(x_{1})+\int \, dx_{1}dx_{2}\, g(x_{1})\star g(x_{2})\star S_{2}(x_{1},x_{2})+....
\ee
$$=1+\sum_{n\geq 1}\frac{1}{n!}\int S_{n}(x_{1},...,x_{n})\star g(x_{1})\star ...\star g(x_{n})dx_{1}...dx_{n}\,.$$

Then, the BS causality condition reads 
\be{deltagx}
\frac{\delta}{\delta g(x)}\Big( \frac{\delta S(g)}{\delta g(y)}\star S^{\dagger}(g)\Big)=0,\,\,\,\qquad x<y,
\ee
implying, from Eq.~\eqref{Smatrix}, the set of identities 
$$C_{n}(y,x_{1},...,x_{n})=0\,,$$
where
\begin{eqnarray}
C_{n}(y,x_{1},...,x_{n})=\!\!\!\!\!\!\!\! && \imath S_{n+1}(y,x_{1},...,x_{n})+ \label{follwing}\\
&&\imath \sum_{0\neq k\neq n-1}P\Big(\frac{x_{1},...,x_{k}}{x_{k+1},...,x_{n}}S_{k+1}(y,x_{1},..,x_{k})\star S_{n-k}^{\dagger}(x_{k+1},...,x_{n}) \Big)\,, \nonumber
\end{eqnarray}
where $P$ is the sum over the distinct ways of partitioning, in $n!/k!(n-k)!$ different ways, a set of elements $\{x_{1},x_{2},...,x_{n}\}$ into two sets of $k$ and $n-k$, namely 
$\{x_{1},x_{2},x_{3},...,x_{k}\}$ and $\{x_{k+1},...,x_{n}\}$. 

The BS causality condition must be used together with the unitarity condition, cast in terms of the start product, and expanded using Eq.~\eqref{Smatrix}, namely
\be{unit}
S[g]\star S^\dagger[g]=1\,.
\ee

Cases $n=1,2$ read as follows: 
\be{cases}
-\imath C_{1}(x,y)=S_{2}(x,y)+S_{1}(x)\star S_{1}^{\dagger}(y)=0\, , 
\ee
\be{Css}
C_{2}(x,y,z)=\imath S_{3}(x,y,z)+\imath S_{1}(x)\star S_{2}^{\dagger}(y,z)+\imath S_{2}(x,y)\star S_{1}^{\dagger}(z)+\imath S_{2}(x,z)\star S_{1}^{\dagger}(y)=0\, . 
\ee

The First condition corresponds to: 
\be{caa}
\imath S_{2}(x,y)+\imath S_{1}(x)\star S_{1}(y)^{\dagger}=\imath S_{2}(x,y)-\imath S_{1}(x)\star S_{1}(y)=0\, , 
\ee
\be{Sdxy}
S_{2}(x,y)=S_{1}(x)\star S_{1}(y)=-T(\mathcal{H}_{\star}(x)\star \mathcal{H}_{\star}(y))\, , 
\ee
which is in turn equivalent to 
\be{condition}
T_{\star}[\mathcal{H}_{1}(x)\star \mathcal{H}_{2}(y)]-T_{\star}[\mathcal{H}_{1}(x)\star \mathcal{H}_{2}(y)]=0, \,\,\,\qquad (x-y)^{2}<0\, . 
\ee

We now specify the Hamiltonian interactions to assume the generic forms
\be{H1H2}
\mathcal{H}_{1}(x)=F_{1}[\phi(x_{1})...\phi(x_{n})]|_{x_{1}=...=x_{n}=x},\,\,\,\qquad \mathcal{H}_{2}(y)=F_{2}[\phi(y_{1})...\phi(y_{n})]|_{y_{1}=...=y_{n}=y}\,.
\ee
We may then evaluate their time ordered commutator to be
\begin{eqnarray}
&&T_{\star}[\mathcal{H}_{1}(x)\mathcal{H}_{2}(y)]-T_{\star}[\mathcal{H}_{1}(x)\mathcal{H}_{2}(y)] \label{O1O2} \\
&&=\theta(y^{0}-x^{0})[O_{1}(x),O_{2}(y)] \nonumber \\
&&-\mathcal{F}_{1}\mathcal{F}_{2}\left\{\theta(Y^{0}-X^{0})[\phi(x_{1})...\phi(x_{n}),\phi(y_{1})...\phi(y_{n}]\right\}\Big|_{x_{1}=...=x_{n}=x,y_{1}=...=y_{n}=y} \,, \nonumber
\end{eqnarray}
$$X^{0}=\frac{\sum_{i=1}^{n}x_{i}^{0}}{n},\qquad Y^{0}=\frac{\sum_{i=1}^{n}y_{i}^{0}}{n}\, , $$
$$\mathcal{F}_{1}=e^{\frac{\imath}{2}\theta^{\mu\nu}(\partial_{\mu}^{x_{1}} \partial_{\nu}^{x_{2}}+...+\partial_{\mu}^{x_{n-1}} \partial_{\nu}^{x_{n}})}\, , $$
$$\mathcal{F}_{2}=e^{\frac{\imath}{2}\theta^{\mu\nu}(\partial_{\mu}^{y_{1}} \partial_{\nu}^{y_{2}}+...+\partial_{\mu}^{y_{n-1}} \partial_{\nu}^{y_{n}})}\, . $$

This proof is enough to conclude that microcausality is violated, contrary to the previous claims in Ref.~\cite{Fiore:2007vg}.

\section{QED}

In this section, we consider the application of the GM deformation to 
standard processes in QED: electron-muon and electron-electron tree level scatterings. 

In GM-QED, the interaction vertex of the charge fermions with the photon fields casts
\be{interactionVert}
H_{I}=e\int d^{4}x \, \bar{\psi}(\hat{x})\star \slashed{A}(\hat{x}^{c})\psi(\hat{x}))\, , 
\ee
\be{psi}
\psi(\hat{x})=\int d\mu(k) \sum_{s}[a^{s}(k)u^{s}(k)e^{-\imath k\cdot \hat{x}}+b^{\dagger(s)}v^{(s)}(k)e^{\imath k\cdot \hat{x}}]\, , 
\ee
\be{psibar}
\bar{\psi}(\hat{x})=\int d\mu(k) \sum_{s}[b^{(s)}\bar{v}^{(s)}(k)e^{- \imath k\cdot \hat{x}}+a^{\dagger(s)}(k)\bar{u}^{(s)}(k)e^{\imath k\cdot \hat{x}}]\, , 
\ee
\be{Agg}
\slashed{A}(\hat{x}^{c})=\int d\mu(k)\sum_{r}[\alpha^{(r)}(k)\slashed{\epsilon}^{(r)}(k)e^{-\imath k\cdot x^{c}}+\alpha^{\dagger (r)}(k)\bar{\slashed{\epsilon}^{(r)}}(k)e^{\imath k\cdot x^{c}}]\, , 
\ee

where 
\be{kakak}
a^{(s_{1})}(p_{1})a^{(s_{2})}(p_{2})=-e^{-\imath p_{1}\wedge p_{2}}a^{(s_{2})}(p_{2})a^{(s_{1})}(p_{1})\,,
\ee
\be{kakaks}
a^{(s_{1})\dagger}(p_{1})a^{(s_{2})\dagger}(p_{2})=-e^{\imath p_{1}\wedge p_{2}}a^{(s_{2})\dagger}(p_{2})a^{(s_{1})\dagger}(p_{1})\,,
\ee
\be{iiiaaii}
a^{(s_{1})}(p_{1})a^{(s_{2})\dagger}(p_{2})=-e^{-\imath p_{1}\wedge p_{2}}a^{(s_{2})\dagger}(p_{2})a^{(s_{1})}(p_{1})+2p_{10}\delta(\vect{p}_{1}-\vect{p}_{2})\, , 
\ee
\be{hkak}
\alpha^{(s_{1})}(p_{1})\alpha^{(s_{2})\dagger}(p_{2})=e^{-\imath p_{1}\wedge p_{2}}\alpha^{(s_{2})\dagger}(p_{2})\alpha^{(s_{1})}(p_{1})+2p_{10}\delta(\vect{p}_{1}-\vect{p}_{2})\, . 
\ee

We start considering an electron-muon scattering process: the in-coming two-particles state reads 
\be{jaka}
|p_{1},s_{1};p_{2}s_{2}\rangle=a^{(s_{1})\dagger}(p_{1}) a^{(s_{2})\dagger}(p_{2}) |0\rangle =e^{\frac{\imath}{2}p_{1}\wedge p_{2}}|0\rangle\,,
\ee
while the out-coming two-particles state is 
\be{kakal}
\langle p_{1}',s_{1}';p_{2}'s_{2}'|=\langle 0| a^{s_{1}'}(p_{1}')a^{s_{2}'}(p_{2}')=e^{-\frac{\imath}{2}p_{1}'\wedge p_{2}'}\langle 0|c^{s_{1}'}(p_{1}') c^{s_{2}'}(p_{2}') \, . 
\ee
Consequently, the amplitude of the electron-muon scattering casts 
\be{nontrivial}
\mathcal{M}_{\theta}=\frac{(-\imath e)^{2}}{2}\int d^{4}x_{1}d^{4}x_{2}\, \theta(x_{10}-x_{20})  \Big(\bar{\psi}_{e}(x_{1})\star (\slashed{A}(x_{1}^{c})\psi_{e}(x_{1}))\Big)\Big(\bar{\psi}_{\mu}(x_{2})\star (\slashed{A}(x_{2}^{c})\psi_{\mu}(x_{2}))\Big)\,,
\ee
which corresponds to 
\be{amplitude}
\mathcal{M}_{\theta}=\frac{e^{2}}{2q^{2}}e^{\frac{\imath}{2}(p_{1}\wedge p_{2}-p_{1}'\wedge p_{2}')}\Big(\bar{u}^{s_{1}'}(p_{1}')\gamma^{\mu}u^{(s_{1})}(p_{1}) \bar{v}^{s_{2}'}(p_{2}')\gamma_{\mu}v^{s_{2}}(p_{2}) \Big) e^{-\frac{\imath}{2}(p_{1}'\wedge p_{1}+p_{2}'\wedge p_{2})}\, . 
\ee

If no helicity is selected, we must average and sum over the spins, following well known rules of standard QED textbooks: 
\be{averaged}
\frac{1}{4}\sum_{spins}|\mathcal{M}|^{2}=\Big|\Phi_{\theta}(p_{1},p_{2},p_{1}',p_{2}')\Big|^{2}\frac{e^{4}}{4q^{4}}{\rm Tr}\Big[(\slashed{p}_{1}'+m_{e})\gamma^{\mu}(\slashed{p}_{1}+m_{e})\Big]
{\rm Tr}\Big[(\slashed{p}_{2}'+m_{\mu})\gamma_{\mu}(\slashed{p}_{2}+m_{\mu})\gamma_{\nu}\Big]\,,
\ee
where 
\be{Phitheta}
\Phi_{\theta}(p_{1},p_{2},p_{1}',p_{2}')=e^{\imath (p_{1}\wedge p_{2}-p_{1}'\wedge p_{2}')}e^{-\frac{\imath}{2}(p_{1}'\wedge p_{1}+p_{2}'\wedge p_{2})}\, . 
\ee
This has a very remarkable implication: since the phase is entering in the modulus, we discover that it disappears completely. Then, a good high-energy approximation is as usual $m_{e}=0$, for which 
\be{approx}
\frac{1}{4}\sum_{spins}|\mathcal{M}|^{2}=\frac{8e^{4}}{q^{4}}\Big|\Phi_{\theta}(p_{1},p_{2},p_{1}',p_{2}')\Big|^{2}\Big[(p_{1}\cdot p_{2}')(p_{1}'\cdot p_{2})+(p_{1}\cdot p_{2})(p_{1}'\cdot p_{2}')-m_{\mu}^{2}(p_{1}\cdot p_{1}')\Big]\, . 
\ee

In the CM of mass system, characterized by 
\be{CM}
p_{1}=(k,k\hat{z}),\,\,\,\quad p_{2}=(E,-k\hat{z}),\,\,\,\quad  p_{1}'=(k,\vect{k}),\,\,\, \quad  p_{2}'=(E,-{\vect{k}})\,,
\ee
one obtains 
\be{CM}
\frac{1}{4}\sum_{spins}|\mathcal{M}|^{2}=\Big|\Phi_{\theta}(E,k,\theta)\Big|^{2}\frac{2e^{4}}{k^{2}(1-\cos \theta)^{2}}\Big( (E+k)^{2}+(E+k\cos \theta)^{2}-m_{\mu}^{2}(1-\cos \theta)\Big)\,,
\ee
where $\phi^{(1),(2)}_{\theta}$ are the $\theta$-phases. 

Continuing our calculations, we consider the total cross section

\be{dsigmadomega}
\Big(\frac{d\sigma}{d\Omega}\Big)_{CM}=\frac{|\mathcal{M}_{\theta}|^{2}}{64\pi^{2}(E+k)^{2}}\,,
\ee

\be{ssigmafinal}
\frac{d\sigma}{d\Omega}=\Big|\Phi_{\theta}(E,k,\theta)\Big|^{2}\frac{\alpha^{2}}{2k^{2}(E+k)^{2}(1-\cos \theta)^{2}}
\Big( (E+k)^{2}+(E+k\cos \theta)^{2}-m_{\mu}^{2}(1-\cos \theta)\Big)\, ,
\ee
where the first exponential phases of the expression come from the deformed creation/annihilation operators, while the last ones arise from the star-product operation. \\

Let us consider the case $\theta_{0i}=0$, for which 
\be{lo}
\log \Phi_{\theta}(E,k,\theta)=\frac{\imath}{2}\Big( (k\hat{z}) \cdot \theta_{ij} \cdot (-k\hat{z})-\vect{k}\cdot \theta \cdot (-\vect{k})-2\vect{k}\cdot \theta \cdot k \hat{z}\Big)=\frac{\imath}{2}\Big( -2\vect{k}\cdot \theta \cdot k \hat{z}\Big)\,,
\ee
where the first two terms are zero, as a contraction of the $\theta$-antisymmetric matrix components with symmetric combinations. Consequently, the only contribution to the standard QED amplitude is a phase, which nonetheless finally disappears in the differential cross section --- see e.g. Eq.~(\ref{ssigmafinal}). This means that the Lorentz symmetry deformation is completely invisible in the CM system. 

At the leading order, this result can be also extended to the electron-proton scattering and, accordingly, to atomic physics. In the case of $\theta_{i0}\neq 0$, we obtain a more complicated but still finally eliding phase: 
\be{Phi}
\log \Phi= \imath \theta_{\hat{z}i}\cdot {\bf k}_{i}k+\imath k^{2}\hat{z}\cdot \theta_{z0}+\imath {\bf k}_{j}\cdot \theta_{j0}k\, . 
\ee
It is worth to note again that just the standard QED result is then obtained, i.e.
\be{ssigmafinal}
\frac{d\sigma}{d\Omega}=\frac{\alpha^{2}}{2k^{2}(E+k)^{2}(1-\cos \theta)^{2}}
\Big( (E+k)^{2}+(E+k\cos \theta)^{2}-m_{\mu}^{2}(1-\cos \theta)\Big)\, . 
\ee
This apodeictically shows that the CPT phases are invisible in the electron-muon scattering. However, the case of electron-electron scattering (or muon/muon, proton/proton scatterings) is less trivial. Considering the s and t channel diagrams, one obtains
\be{Mtheta}
\mathcal{A}_{\theta}=e^{\imath(p_{1}\wedge p_{2}-p_{1}'\wedge p_{2}')}(\mathcal{A}_{1}e^{-\imath\delta}-\mathcal{A}_{2}e^{\imath\delta})\,,
\ee
where 
\be{delta}
\delta=\frac{1}{2}(p_{1}'\wedge p_{1}+p_{2}'\wedge p_{2})\,,
\ee
and 
\be{MuMd}
|\mathcal{A}_{\theta}|^{2}=|\mathcal{A}_{1}|^{2}+|\mathcal{A}_{2}|^{2}+\mathcal{A}_{1}^{*}\mathcal{A}_{2}e^{+2\imath \delta}+\mathcal{A}_{2}^{*}\mathcal{A}_{1}e^{-2\imath\delta}\, . 
\ee

In the CM, for $\theta_{0i}=0$, the phase reads 
\be{lambda}
\delta={\bf k}_{i}\cdot \theta_{i\hat{z}}k\,,
\ee

while, for $\theta_{0i}\neq 0$, one finds 
\be{lambda}
\delta={\bf k}_{i}\cdot \theta_{i\hat{z}}k+k\theta_{0j}\cdot  k\hat{z}+{\bf k}\cdot \theta_{i0}k+E\theta_{0j}\cdot (-k\hat{z})+(-{\bf k}\cdot \theta_{i0}E)\,.
\ee

In the massless limit, one then obtains 

\be{masslesslimit}
\langle |\mathcal{A}|^{2}\rangle=2e^{4}\Big( \frac{s_{14}^{2}}{s_{13}^{2}}+\frac{s_{13}^{2}}{s_{14}^{2}}+\frac{s_{12}^{4}}{s_{13}^{2}s_{14}^{2}}\cos (2\delta)     \Big)=2e^{4}\Big( \frac{u^{2}}{t^{2}}+\frac{t^{2}}{u^{2}}+\frac{s^{4}}{t^{2}u^{2}}\cos (2\delta)  \Big) \,, 
\ee

which in the CM is 
$$\frac{e^{4}}{4E^{4}}f(\theta,\delta)\,,$$

where 
$$f(\theta,\delta)=\frac{1+c^{4}(\theta/2)}{s^{4}(\theta/2)}+\frac{1+s^{4}(\theta/2)}{c^{4}(\theta/2)}+\frac{4}{s^{2}(\theta/2)c^{2}(\theta/2)}\cos \delta \, . $$

The cross section now reads
\be{cross}
\frac{d\sigma}{d\Omega}=\Big(\frac{e^{2}}{4\pi}\Big)^{2}\frac{1}{8E^{2}}f(\theta,\delta)\, . 
\ee

Also in this case, CPT invariance seems to be preserved. This is because the $\theta$-term is inside an even function. However, it still leaves an indirect characteristic imprinting 
introducing a cosinusoidal modulation $\cos \delta$ overimposed on the standard QED result. 
This is an effect of the violation of the Spin-Statistic theorem and of the 
violation of Fermi-Dirac statistical indinstinguibility of s-channels and t-channels. 



\section{Violation of Pauli in multi-particle states}

In this section, we elaborate on Pauli Exclusion Principle violations in 
GM-Standard Model. 
Let us consider the one-particle state in the form  
\be{alpha}
|\alpha\rangle =\langle a^{\dagger},\alpha |0\rangle=\langle c^{\dagger},\alpha |0\rangle=\int \frac{d^{d}p}{2p_{0}}\alpha(p)c^{\dagger}(p)\, ,
\ee
where 
\be{alpha}
\langle\alpha|\alpha\rangle=1,\,\,\,\qquad  \int \frac{d^{d}p}{2p_{0}}|\alpha(p)|^{2}=1\, . 
\ee
From the definition in Eq.~(\ref{alpha}), we may construct a two-identical-particles state that reads  
\be{two}
|\alpha,\alpha\rangle = \langle a^{\dagger},\alpha\rangle  \langle a^{\dagger},\alpha\rangle  |0\rangle\, =
\ee
$$=\int \frac{d^{d}p_{1}}{2p_{10}}\frac{d^{d}p_{2}}{2p_{10}}e^{-\frac{\imath}{2}p_{1\mu}\theta^{\mu\nu} p_{2\nu}}\alpha(p_{1}) \alpha(p_{2})c^{\dagger}(p_{1})c^{\dagger}(p_{2})|0\rangle\,.$$
This state must be normalized considering that its norm is expressed by 
\be{alphaapha}
N=\langle\alpha,\alpha|\alpha,\alpha \rangle=\int \frac{d^{d}p_{1}}{2p_{10}}\frac{d^{d}p_{2}}{2p_{20}}(\bar{\alpha}(p_{1})\alpha(p_{1}))(\bar{\alpha}(p_{2})\alpha(p_{2}))(1-e^{-\imath p_{1\mu}\theta^{\mu\nu} p_{2\nu}})
\ee
$$
=\int \frac{d^{d}p_{1}}{2p_{10}}\frac{d^{d}p_{2}}{2p_{20}}(\bar{\alpha}(p_{1})\alpha(p_{1}))(\bar{\alpha}(p_{2})\alpha(p_{2}))(1-\cos( p_{1\mu}\theta^{\mu\nu} p_{2\nu}))\,,$$
since the $\sin$-part is odd under the interchange of $p_{1}\leftrightarrow p_{2}$ 
and null in the integral. Consequently, we can redefine the two-particles state as a normalized one, which reads 
\be{norm}
|\alpha,\alpha\rangle \rightarrow  \frac{1}{N(\alpha,\alpha)}|\alpha,\alpha\rangle,\,\,\qquad \, \langle \alpha|\alpha\rangle=1\, .
\ee

Now let us calculate the transition amplitude for the overlap probability that a two-different-particles state evolves into a two-identical-particles state. In the case of fermions, this casts
\be{alphalpa}
\langle \beta,\gamma|\alpha, \alpha\rangle=\frac{1}{N}\int \frac{d^{d}p_{1}}{p_{10}}\frac{d^{d}p_{2}}{p_{20}}(\bar{\beta}(p_{1})\alpha(p_{1}))(\bar{\gamma}(p_{2})\alpha(p_{2}))\Big[1-e^{-\imath p_{1\mu}\theta^{\mu\nu} p_{2\nu}} \Big]
\ee
$$= \frac{1}{N}\int \frac{d^{d}p_{1}}{p_{10}}\frac{d^{d}p_{2}}{p_{20}}(\bar{\beta}(p_{1})\alpha(p_{1}))(\bar{\gamma}(p_{2})\alpha(p_{2}))\Big[1-\cos\Big(p_{1\mu}\theta^{\mu\nu} p_{2\nu}\Big) \Big]\,.$$

For $\theta\rightarrow 0$, the overlap amplitude vanishes. However, for $\theta \neq 0$, the Pauli principle can be violated if the states are composed of fermions. In other words, a two-fermions state can transit into a state in which fermions are identical. Not only a GM cosinusoidal phase would then appear, but also other phases provided by the electromagnetic interactions show up. Let us consider indeed the GM effective Hamiltonian density, which is expressed by  
\be{HHH}
H_{GM,ij}=\langle \Psi_{i}^{\theta}|V_{\theta}|\Psi_{i}^{\theta}\rangle=\langle \Psi_{i}^{0}|\mathcal{H}_{E}|\Psi_{j}^{0}\rangle = V_{0}\Big\{\cos (\phi_{PEPV})-\cos\Big(\phi+p_{1\mu}\theta^{\mu\nu} p_{2\nu}\Big)\Big\}\,,
\ee
where, for a central potential, 
\be{central}
2\phi_{PEPV}=p_{1}\wedge p_{2}-p_{1}'\wedge p_{2}'-p_{1}'\wedge p_{1}+p_{2}'\wedge p_{2}\,.
\ee
Here the phases are provided both by Eq.~\eqref{alphalpa} and the central interaction potential. 

\subsection*{Atomic levels transitions}

Now, let us focus on the specific problem of atomic level transitions.
In this case, we can consider a non-relativistic limit approach based on 
perturbation theory. The effective Hamiltonian is the 0th order standard one 
plus a PEPV term:  $H=H_{0}+V_{I,0}+V_{I,0}\phi_{PEPV}^{2}$, where 
$\phi_{PEPV}$ is the PEPV GM energy dependent phase. 
The 1st order perturbation coefficient acquires the form 
\be{ckkk}
\dot{c}_{b}^{1}(t)=(\imath \hbar)^{-1}H_{ba}'(t)e^{\imath \omega_{ba}t}\,,
\ee
and if the perturbation is time independent, recasts 
\be{cbb}
c_{b}^{(1)}(t)=-\frac{H'_{ba}}{\hbar \omega_{ba}}(e^{\imath \omega_{ba}t}-1). 
\ee
The transition probability is then found to be 
\be{PPP}
P_{ba}(t)=|c_{b}^{(1)}(t)|^{2}=\frac{2}{\hbar}|H'|^{2}F(t,\omega_{ba})=\frac{2}{\hbar}V_{0}^{2}\phi^{2}F(t,\omega_{ba})\,,
\ee
with $F=(1-\cos \omega t)/\omega^{2}$. Operationally, within the long-time limit, this expression leads to $F\rightarrow \pi t \delta(\omega)$, and then finally we recover 

\be{WWW}
W=\frac{2\pi}{\hbar}|H'_{ba}|^{2}=\frac{2}{\pi \hbar}V_{0}^{2}\phi^{2}_{PEPV}=W_{0}\phi^{2}_{PEPV}\,.
\ee
The expression encodes a suppression that is quartic in the ratio  between $\sqrt{\theta}$ and energy-momentum components. However, the suppression may be an artifact of the number of fields involved in the initial state. Indeed, if one considers Eq.~(\ref{alphalpa}) in presence of three particles, odd despite of even powers may be suppressed in the integral, leading to a linear order corrections in the phase $\phi_{PEPV}$. In this latter case we obtain
\be{WWW}
W\simeq W_{0}\phi_{PEPV}\, ,
\ee
where 
$\phi_{PEPV}=\delta^{2}$ can be falsified from experimental measures.

\subsection*{Observable quantities}

We devote the last part of this section to the already developed experimental set-up that can be deployed to falsify proposals within the  the GM-Standard Model. We may distinguish two cases, corresponding to different choices of the $\theta$-components. \\

For the first choice,   
\be{thetazi}
\theta_{0i}=0\rightarrow \phi_{PEPV}=\frac{1}{2}\Big(p_{1}^{i}\theta_{ij} p_{2}^{j}-p_{1}'^{i}\theta_{ij} p_{2}'^{j}-p_{1}'^{i}\theta_{ij} p_{1}^{j}+p_{2}'^{i}\theta_{ij} p_{2}^{j}\Big)\,.
\ee
Considering the particle $1$ as an electron and the particle $2$ as a nucleus, all terms involving $p_{1}$ and $p_{1}'$ are subleading. On the other hand $|p_{2}|$ and $|p_{2}'|$ are of the order of the energy levels in the atom. For this set-up, the GM-Standard Model predicts the result
\be{EEEE}
\phi_{PEPV}\simeq \frac{1}{2} C \frac{\bar{E}_{1}}{\Lambda}\frac{\bar{E}_{2}}{\Lambda}\,,
\ee
where $\bar{E}_{1,2}$ are the energy levels occupied by the initial and the final electrons, the quantity $C$ reads $C=\hat{\bar{{\bf p}}}_{1}\cdot \theta \cdot \hat{{\bf p}}_{2}$, and we introduce the UV cutoff scale $\Lambda$, which is hidden in the definition of $\theta$.

For the second choice, 
\be{thetazi}
\theta_{0i}\neq 0\rightarrow \Delta \phi_{PEPV}=\frac{1}{2}\Big(p_{1}^{0}\theta_{0j} p_{2}^{j}-p_{1}'^{0}\theta_{0j} p_{2}'^{j}-p_{1}'^{0}\theta_{0j} p_{1}^{j}+p_{2}'^{0}\theta_{0j} p_{2}^{j}\Big)+(0\leftrightarrow j)\,,
\ee
with
\be{PEPVphi}
\phi_{PEPV}\simeq  \frac{D}{2}\frac{E_{N}}{\Lambda}\frac{\Delta E}{\Lambda}   \, , 
\ee
where $E_{N}\simeq m_{N}\simeq A m_{p}$ is the nuclear energy, and $\Delta E=E_{2}-E_{1}$ is the transition energy of the electron.

Concretely, this scenario can be tested by the VIP experiments, searching for PEPV transitions $2s \rightarrow 1s$ in Cooper atoms \cite{Pichler:2016xqc}. In the case of VIP, $\bar{E}_{1}=\bar{E}_{2p_{1/2}}$ and $\bar{E}_{2}=\bar{E}_{1s_{1/2}}$ with a transition energy of $7\, KeV$.

In a similar non-relativistic approach, we can also estimate 
nuclear Pauli violating transitions. 
For radiation emitting nuclear transitions, we obtain: 
\be{EEEE}
\phi_{PEPV}\simeq \frac{1}{2} C \frac{\bar{E}_{1}}{\Lambda}\frac{\bar{E}_{2}}{\Lambda}\,.
\ee

When transitions emitting a proton or a neutron are taken into account, we find again the expression 
\be{EEEE}
\phi_{PEPV}\simeq \frac{1}{2} C \frac{\bar{E}_{1}}{\Lambda}\frac{\bar{E}_{2}}{\Lambda}\,,
\ee
but in which now $E_{2}$ corresponds to the critical energy for which proton is unbounded, despite of the energy level transition. For example, in DAMA experiments the characteristic of the experiments are such that $\bar{E}_{2}\simeq 10\, {\rm MeV}$ \cite{Bernabei:2009zzb}.

\section{Radiative corrections and IR/UV mixings}

In this section, we discuss the still ``obscure'' phenomenon of the IR/UV mixing. 
A general one-loop diagram with $N$ vertices in a $\phi^{n}$ theory reads \be{AAA}
\int d^{d}\phi(x)_{*}^{n}=\int \frac{d^{d}k_{1}}{(2\pi)^{n}}...\frac{d^{d}k_{n}}{(2\pi)^{n}}(2\pi)^{d}\delta\Big( \sum_{j=1}^{n}k_{j}\Big)
\phi(k_{1})....\phi(k_{n})e^{-\frac{\imath}{2}\sum_{i<j}k_{i}\wedge k_{j}}
\ee
Because of the presence of phases, planar and non-planar Feynman diagrams will entail different contributions. Let us focus on the simplest case, the $\lambda \phi_{\star}^{4}$ theory, and consider the mass-renormalization self-energy diagram. Within the standard theory only the contribution from one diagram shall be calculated. Nonetheless, in the non-commutative theory another quadratically divergent non-planar contribution arises, the expression of which reads
\be{Pip}
\Pi(p)_{\rm non-planar}=\frac{\lambda}{6}\int \frac{d^{4}k}{(2\pi)^{4}}\frac{e^{\imath k \wedge p }}{k^{2}+m^{2}}\, .
\ee
This divergent contribution corresponds to have an energy dependent effective cutoff of the form
\be{Lambdaeff}
\Lambda_{\rm eff}^{-2}=\Lambda^{-2}+\tilde{p}^{2}\, 
\ee
with $\tilde{p}^\mu=\theta^{\mu\nu} p_\nu$. For $\tilde{p}\rightarrow 0$, Eq.~\eqref{Lambdaeff} entails $\Lambda\rightarrow \infty$. Again, we are in disagreement with Ref.~\cite{Fiore:2007vg}, in which it has been claimed that no UV/IR mixings appear in the radiative corrections. 

\subsection*{QED and IR/UV mixing}

Let us focus now non-commutative QED, taken into account without matter: 
\be{SSS}
S=-\frac{1}{4g^{2}}\int d^{4}x\, F_{\mu\nu}F^{\mu\nu}\,,
\ee
\be{FmuNU}
F_{\mu\nu}=\partial_{\mu}A_{\nu}-\partial_{\nu}A_{\mu}-\imath(A_{\mu}\star A_{\nu}-A_{\nu}\star A_{\mu})\, . 
\ee
It is worth to note that non-commutativity induces new photon self-interaction vertices. In particular, the three photon vertex $(p_{1},\mu),(p_{2},\nu),(p_{3},\sigma)$ reads 
\be{vertex}
2g \sin\Big(\frac{1}{2}p_{1}\wedge p_{2} \Big)[\eta^{\mu\sigma}(p_{1}-p_{3})^{\nu}+{\rm perm.}]\, ,
\ee
while the four photon vertex $(p_{1},\mu),(p_{2},\nu),(p_{4},\rho),(p_{3},\sigma)$ recasts 
\be{ajjaj}
-4\imath g^{2}\Big[\sin\Big(\frac{1}{2}p_{1}\wedge p_{2}   \Big)\sin\Big(\frac{1}{2}\tilde{p}_{4}\cdot p_{3} \Big)(\eta^{\mu\rho}\eta^{\nu\sigma}-\eta^{\mu\sigma} \eta^{\nu\rho}) +{\rm perm.} \Big]\, . 
\ee
Finally, in a BRST fashion, the ghosts-photons vertex reads 
\be{ghostphoton}
2g\, p_{2}^{\mu}\sin\Big(\frac{1}{2}p_{1}\wedge p_{2} \Big)\, . 
\ee
Indeed, in the BRST quantization scheme one may find
\be{ZHHZ}
Z[J,\bar{\eta},\eta]=\int \!\! \mathcal{D}A\, \mathcal{D}b\, \mathcal{D}\bar{c}\, \mathcal{D}c\,\, e^{ \imath S+\imath \int \!\!d^{4}x\, (JA +\bar{\eta}c+\bar{c}\eta   )}\,,
\ee
where $b$ is the auxiliary field that instantiates the Lorentz condition $\partial A=f$, with $f$ an arbitrary function of $x$, and $\bar{c}$ and $c$ are ghost fields. The gauge-fixing classical action then casts 
\be{SS}
S=\int d^{4}x \Big( -\frac{1}{4g^{2}}F^{\mu\nu}\star F_{\mu\nu}+\frac{\alpha}{2}g^{2}b\star b-b\star \partial A+\bar{c}\star \partial_{\mu} D^{\mu} c  \Big)\,.
\ee
From Eq.~\eqref{SS}, we can calculate the vacuum polarization diagram that arises from the photon self-interactions, namely  
\be{Pimunna}
\imath \Pi_{\mu\nu}(p)=\frac{2\imath}{\pi^{2}}\frac{\tilde{p}_{\mu}\tilde{p}_{\nu}}{(p\,o\,p)^{2}}-\frac{\imath}{16\pi^{2}}\Big(\frac{13}{3}-\alpha \Big)\Big[\frac{1}{\epsilon}-\log(p^{2}\,p\, o\, p)   \Big](p^{2}\eta^{\mu\nu}-p^{\mu}p^{\nu})\, .
\ee
In Eq.~\eqref{Pimunna} we have introduced the positive-definite inner product 
$$p\, o \, q =-p_\mu(\theta^2)^{\mu \nu}q_\nu= q\, o\, p\,,$$
with $(\theta^2)^{\mu \nu}=\delta_{\lambda\rho} \theta^{\mu\lambda} \theta^{\rho\nu}$, and the vertex correction 
$$\imath\Gamma_{\mu_{1}\mu_{2}\mu_{3}}=\frac{2}{\pi^{2}}\cos\Big(\frac{p_{1}\wedge p_{2}}{2} \Big)
 \sum_{i}^{3}\frac{(\tilde{p}_{i})_{\mu_{1}}(\tilde{p}_{i})_{\mu_{2}}(\tilde{p}_{i})_{\mu_{3}}}{(p_{i}\,o\,p_{i})^{2}}$$
 $$+\frac{1}{16\pi^{2}}\sin\Big(\frac{p_{1}\wedge p_{2}}{2}\Big)\Big(\frac{17}{3}-3\alpha \Big)\Big[\frac{1}{\epsilon}-\frac{1}{3}\sum_{i} \log(p_{i}\, o\,p_{i})  \Big]$$
\be{Gammama}
\times [\eta_{\mu_{1}\mu_{2}} (p_{1}-p_{2})_{\mu_{3}}+\eta_{\mu_{2}\mu_{3}} (p_{2}-p_{3})_{\mu_{1}}+\eta_{\mu_{3}\mu_{1}} (p_{3}-p_{1})_{\mu_{2}} ]\,.
\ee
This entails an appearance of the UV/IR mixing in QED. The one-loop correction to the photon dispersion relation then leads to a pathological ghost-like behaviour, 
\be{omea}
\omega^{2}=p^{2}-\frac{2g^{2}}{\pi}\frac{1}{\theta^{2}(p_{1}^{2}+p_{2}^{2})}\,.
\ee
It is still unknown whether the divergences that occur in non-commutative QED shall be still considered as really existing pathologies of the theory, or whether the IR radiation, or the Cherenkov like radiation, may cure these latter. IR/UV mixing thus remains an open question within the GM-Standard Model. 





\section{Connections with string theory}

In this section we review the relations among string theory and the non-commutative $\theta$-Poincar\'e QFT models, as proposed by Seiberg and Witten in Ref.~\cite{Seiberg:1999vs}. The bosonic string sector can be recast as \be{actionn}
S=\frac{1}{4\pi\alpha'}\int_{\Sigma}(g_{ij}\partial_{a}x^{i}\partial^{a}x^{j}-2\pi \imath \alpha' B_{ij}\epsilon^{ab}\partial_{a}x^{i}\partial_{b}x^{j})
\ee
$$=\frac{1}{4\pi \alpha'}\int_{\Sigma}g_{ij}\partial_{a}x^{i}\partial^{a}x^{j}-\frac{\imath}{2}\int_{\partial \Sigma}B_{ij}x^{i}\partial_{t}x^{j}\,,$$
where $\Sigma$ is the string worldsheet, $i$ are the space-time coordinates and $a$ denote the internal worldsheet coordinates, while $B$ stands for the antisymmetric form and $X$ for the bosonic string fields. 

The Boundary condition on Dp-branes are
\be{gahjakl}
g_{ij}\partial_{n}x^{j}+2\pi \imath \alpha' B_{ij}\partial_{t}x^{j}|_{\partial \Sigma}=0\,,
\ee
$\partial_{n}$ denoting derivatives along the normal direction to $\partial \Sigma$. 

We imagine $\Sigma$ as a disk, a typical approximation for open strings theory. In this case, one obtains the condition
\be{gijk}
g_{ij}(\partial-\bar{\partial})x^{j}+2\pi \alpha' B_{ij}(\partial+\bar{\partial})x^{j}|_{z=\bar{z}}=0\,,
\ee
where $\partial=\partial/\partial z$ and $\bar{\partial}=\partial/\partial \bar{z}$. 

In such a theory, when these boundaries are taken into account, the propagator reads 
\be{GAJkl}
\langle x^{i}(z) x^{j}(z')\rangle=-\alpha' \Big[ g^{ij}\log|z-z'|-g^{ij}\log|z-\bar{z}'|
\ee
$$+G^{ij}\log|z-\bar{z}'|^{2}+\frac{1}{2\pi \alpha'}\theta^{ij}\log\frac{z-\bar{z}'}{\bar{z}-z'}+D^{ij}\Big]\, ,$$
where 
\be{Gkopa}
G^{ij}=\Big( \frac{1}{g+2\pi \alpha' B}\Big)^{ij}_{S}=\Big(\frac{1}{g+2\pi \alpha'B}g\frac{1}{g-2\pi \alpha'B}\Big)\, , 
\ee
\be{Gijk}
G_{ij}=g_{ij}-(2\pi \alpha')^{2}(Bg^{-1}B)_{ij}\, , 
\ee
\be{theta}
\theta^{ij}=2\pi \alpha'\Big(\frac{1}{g+2\pi \alpha' B} \Big)_{A}^{ij}=-(2\pi \alpha')^{2}\Big(\frac{1}{g+2\pi \alpha' B} B \frac{1}{g-2\pi \alpha' B}\Big)\,,
\ee
with $(...)_{S,A}$ denoting the symmetric and antisymmetric parts of the matrices. 

On the boundary points, the propagator simplifies to 
\be{alap}
\langle x^{i}(\tau)x^{j}(\tau')\rangle=-\alpha' G^{ij}\log( \tau-\tau')^{2}+\frac{\imath}{2}\theta^{ij}\epsilon(\tau-\tau')\,,
\ee
where $\epsilon=\pm 1$ denotes positive or negative values of $\tau$. Interpreting $\tau$ as the time, we then find 
\be{ghjk}
[x^{i}(\tau),x^{j}(\tau)]=T(x^{i}(\tau)x^{j}(\tau^{-}))-T(x^{i}(\tau)x^{j}(\tau^{+}))= \imath \theta_{ij}\,.
\ee

The tachyon vertex reads 
\be{thjk}
e^{\imath p\cdot x}(x)\cdot e^{\imath q\cdot x}(\tau')\sim (\tau-\tau') e^{2\alpha'G^{ij}p_{i}q_{j}}e^{-\frac{\imath}{2}\theta^{ij}p_{i}q_{j}}e^{\imath (p+q)\cdot x}(\tau')\,.
\ee
When ignoring $(\tau-\tau')$, this would collapse into
\be{eeee}
\longrightarrow \quad  e^{\imath p\cdot x}\star e^{\imath p\cdot x}(\tau')\,. 
\ee

Focusing now on the limit $\alpha'\rightarrow 0$, one finds
\be{GGGGAK}
G^{ij}=-\frac{1}{(2\pi \alpha')^{2}}\Big(\frac{1}{B}g\frac{1}{B}\Big)^{ij},\,\,\,\qquad  i,j=1,...,r
\ee
\be{oeaa}
G^{ij}=g^{ij},\,\,\, \qquad {\rm otherwise}
\ee
\be{hjklp}
\theta^{ij}=(1/B)^{ij}\qquad  i,j=1,...,r\,,
\ee
while $0$ otherwise. 

In this limit, the propagator becomes non-commutative, i.e.
\be{jklp}
\langle x^{i}(\tau)x^{j}(0)\rangle=\frac{\imath}{2}\theta^{ij}\epsilon(\tau)\,.
\ee

The propagator's structure implies that the normal ordered operators have the properties 
\be{jklaaa}
:e^{\imath p_{i}x^{i}(\tau)}::e^{\imath q_{i}x^{i}(0)}:=e^{-\frac{1}{2}\theta^{ij}p_{i}q_{j}\epsilon(\tau)} :e^{\imath px(\tau)+\imath qx(0)}:\, , 
\ee
while for generic functionals of the string fields one finds 
\be{kll}
:f(x(\tau))::g(x(0)):=e^{\frac{\imath}{2}\epsilon(\tau)\theta^{ij}\frac{\partial}{\partial x^{i}(\tau)}\frac{\partial}{\partial x^{j}(0)}}f(x(\tau))g(x(0))\,.
\ee

\noindent 

\section{Phenomenology of $\theta$-Poincar\'e: Pauli forbidden transitions in underground experiments } \label{mico}

\noindent


\noindent 
As shown in the previous sections, Pauli-Exclusion-Principle violating transitions are unavoidably predicted by the GM-Standard Model. The $\phi_{PEPV}$ phase corresponds to the $\delta^{2}$ parameter measured in the experiments. In Fig.1-2 of Ref.\cite{Addazi:2017bbg}, we showed limits on the relative strength ($\delta^2$) parameter for the searches of new non-paulian transitions. Several methodologies of experimental investigations have been hitherto proposed. 

\subsection*{Atomic transitions}

The VIP experiment \cite{Pichler:2016xqc} searches for PEP forbidden atomic transitions in copper. Its experimental technique consists in the injection of ``fresh" electrons into a copper strip, by means of a circulating current, and in the search for the X-rays which are PEP forbidden radiative transitions --- related to the electrons captured by a copper atom and cascades down to the (already-filled) 1S wave state. The VIP experiment is searching for the $K_\alpha$ (2P $\rightarrow$ 1S) exotic transition. The energy of this PEP forbidden transition is $7.729\, {\rm keV}$, and it would differ from the ordinary $K_\alpha$ transition energy ($8.040\, {\rm keV}$) by a energy difference $\Delta E$ term (about $300\,{\rm eV}$) --- due to the presence line of the other electrons in the already-filled 1S shell. Such a $\Delta E$ shift can be detected by high resolution CCD devices.\\

PEP forbidden electromagnetic atomic transitions can be also searched for in Iodine atoms deploying NaI(Tl) detectors, as done in the DAMA/LIBRA  \cite{Bernabei:2009zzb} and the ELEGANTS V \cite{Ejiri:1992} experiments. Other studies of PEPV electromagnetic transitions in Germanium atoms in PPC HPGe detectors were shown by the MALBEK experiment \cite{Abgrall:2016wtk}. In these cases, PEPV electronic transitions emit X-rays and Auger electrons, directly by the transition itself and later as the following rearrangements of the atomic shell. Let us remark that the detection efficiency of such radiation in the NaI(Tl) detectors of DAMA/LIBRA is almost $100\%$, at the low energy of the process.
This implies that all the ionization energy, for the considered shell, can be detected, shifted by a $\Delta E$ --- due to the presence of the other electrons in filled shells. In this class of experiments, the K-shell is the one considered since providing the largest (available) energy of X-rays /Auger-electrons radiation emissions. Other stringent limits are also achieved by DAMA/NaI on transitions to L-shell, in Iodine atoms \cite{bernabei2}, with $4\div 5$ keV radiation emission --- thanks to the low energy thresholds of NaI(Tl) detectors.

The most stringent constraint on PEPV in atomic transitions comes from the DAMA/LIBRA experiment:
searching for PEPV K-shell transitions in Iodine, with data corresponding to 0.53 ton$\times$yr,  
a lower limit on the PEPV transition lifetime of $4.7 \times 10^{30}$ s has been reached.
This provides a limit on the PEPV phase of 
  $\phi_{PEPV}=\delta^2 < 1.28 \times 10^{-47}$ at 90\% C.L. \cite{Bernabei:2009zzb}. This entails very strong constraints on the non-commutativity scale:
in the magnetic-like $\theta$-Poincar\'e $\Lambda < 10^{18}\, {\rm GeV}$, while in the electric-like phase we obtain from Eq.~(\ref{PEPVphi}) that the limit is less stringent, i.e. $\Lambda <5\times 10^{16}\, {\rm GeV}$.

\subsection*{Nuclear transitions}

From DAMA/LIBRA collaboration also limits on PEPV nuclear transitions can be achieved \cite{Bernabei:2009zzb}. PEPV transitions in nuclear shells of $^{23}$Na and $^{127}$I are investigated, by analyzing possible protons emitted with an energy of E$_p \ge$ 10 MeV. 
The rate of emission of high energy protons (E$_p \ge$ 10 MeV) due to PEPV transitions in $^{23}$Na and $^{127}$I was constrained to $\lsim 1.63 \times 10^{-33}$ s$^{-1}$ (90\% C.L.) \cite{Bernabei:2009zzb},  
corresponding to $\delta^2 \lsim 4 \times 10^{-55}$ at 90\% C.L. In both the electric and the magnetic like $\theta$-Poincar\'e, events with a non-commutative scale that may reach the Planck scale energy are largely excluded.

PEPV has also been tested within the Borexino experiment \cite{Bellini:2010}, analyzing nuclear transitions in the $^{12}$C nuclei. Borexino has an extremely low background and large mass (278 tons) detector, rendering its sensitivity extremely good. Borexino searched for $\gamma$, $\beta^\pm$, neutrons, and protons, emitted in a PEPV transition of nucleons from the 1P$_{3/2}$ shell to the already filled 1S$_{1/2}$ shell. The  following limits on the lifetimes for the different PEP violating transitions were set \cite{Bellini:2010}
(all the limits are at 90\% C.L.):
$\tau$($^{12}$C $\rightarrow ^{12}\widetilde{C} + \gamma$) $\ge 5.0 \times 10^{31}$ yr;
$\tau$($^{12}$C $\rightarrow ^{12}\widetilde{N} + $e$^{-} + \bar{\nu}_e$) $\ge 3.1 \times 10^{30}$ yr; 
$\tau$($^{12}$C $\rightarrow ^{12}\widetilde{B} + $e$^{+}  + \nu_e$)       $\ge 2.1 \times 10^{30}$ yr; 
$\tau$($^{12}$C $\rightarrow ^{11}\widetilde{B} + $p)      $\ge 8.9 \times 10^{29}$ yr and
$\tau$($^{12}$C $\rightarrow ^{11}\widetilde{C} + $n)      $\ge 3.4 \times 10^{30}$ yr.
These limits correspond to PEPV phases from the electromagnetic, strong and weak transitions, as follows:
$\delta^2_\gamma \le 2.2 \times 10^{-57}$,
$\delta^2_N      \le 4.1 \times 10^{-60}$, and
$\delta^2_\beta  \le 2.1 \times 10^{-35}$ \cite{Bellini:2010}.
In both the electric and the magnetic like $\theta$-Poincar\'e, 
$\delta_{\gamma,N}^{2}$ limits largely exclude a non-commutative scale, up to the Planck scale energy.

Other constrains on nuclear transitions were provided by the Kamiokande experiment, consisting on a large underground water Cherenkov detector \cite{Suzuki:1993zp} searching for PEPV emission of $\gamma$ rays in the energy range $19 - 50$ MeV. Also from the Kamiokande experiment, no statistically significant excess was found above the experimental background.
Kamiokande results allow to set a limit on the lifetime of PEPV transitions to $9.0 \times 10^{30} \times$ Br($\gamma$) yr per oxygen nucleus --- where Br($\gamma$) is the branching ratio of the $^{16}$O decays in the $\gamma$ channel. From Kamiokande, PEPV transitions that are due to the p-shell nucleons can be limited to $1.0 \times 10^{32} \times$ Br($\gamma$) yr, corresponding to 
$\delta^2 < 2.3 \times 10^{-57}\, (90\% C.L.)$  \cite{Suzuki:1993zp}. Also this limit, arising from Kamiokande, rules out both the magnetic and the electric-like $\theta$-non-commutativity.




\vspace{0.5cm} 

{\large \bf Acknowledgments} 

\vspace{0.5cm}

We thank Rita Bernabei, Pierluigi Belli,  Zurab Berezhiani, Giovanni Amelino-Camelia, Catalina Oana Curceanu, Giampiero Mangano,
Kristian Piscicchia and Yong Shi-Wu for useful discussions and remarks on this subject. 
We acknowledge support by the NSFC, through the grant No. 11875113, the Shanghai Municipality, 
through the grant No. KBH1512299, and by Fudan University, through the grant No. JJH1512105.





\end{document}